# Equation of the Aftershocks and its Phase Portrait

## A.V. Guglielmi

*Schmidt Institute of Physics of the Earth, Russian Academy of Sciences, Moscow, Russia,*
[guglielmi@mail.ru](mailto:guglielmi@mail.ru)

**Abstract**

Recently, simple differential models of aftershocks have attracted increased attention of researchers at the Institute of Physics of the Earth, Russian Academy of Sciences for their remarkable properties. The shortened Bernoulli equation allows us to introduce the concept of the source deactivation factor and phenomenologically describe the nonstationary of the geological environment in the earthquake source, "cooling down" after the main shock. The Verhulst logistic equation makes it possible to take into account the transition of the aftershock flow to the background seismicity mode. The Kolmogorov-Petrovsky-Piskunov nonlinear diffusion equation is convenient for modeling the spatio-temporal evolution of aftershocks. This paper focuses on the phase portrait of a dynamical system that simulates the evolution of aftershocks. The phase portrait allows us to understand why the Omori hyperbolic law is most clearly manifested after strong earthquakes. It is noted that the Omori epoch, during which the deactivation coefficient is constant, does not immediately end with a transition to the background mode. In the experience after the end of the Omori epoch, the deactivation coefficient changes over time in a rather complex way, and can even becomes negative. The theory predicts that if negative values of the coefficient are held long enough, then a geotectonic explosion occurs, vaguely resembling the main shock of earthquake. The paper discusses possible directions for the further development of the theory.

## 1. Introduction

The problems of the physics of earthquakes are complex and diverse. The relevant literature is extensive and rich in meaningful results. This paper focuses on aftershocks that occur after the main shock of an earthquake. The choice of this topic is mainly related to the scientific interests of the author, who over the past years has studied aftershocks together with his colleagues in the framework of several scientific projects (see, for example, reviews [1–3]). Our approach is based on the phenomenological equations describing the evolution of aftershocks.

Many papers, reviews and monographs talk about two empirical laws of aftershock physics, namely, Omori's law and Bath's law. The Omori law states that the frequency of aftershocks $n(t)$, on average, hyperbolically decreases with time [4]. The Bath law is formulated as follows: the minimum difference between the magnitude of the main shock and the magnitude of the strongest aftershock exceeds a value somewhat larger than unity [5]. (Usually indicate the value of $\triangle M = 1.2$ .) Both laws are widely used in seismology. In this paper, we want to draw attention to interesting prospects for further study of the Omori law in the phase space of a simple dynamic system that simulates an earthquake source.

## 2. Phase portrait

It is convenient to represent the Omori law [4] in the form of a one-parameter evolution equation describing the averaged dynamics of aftershocks [6]

$$\dot{n} + \sigma n^2 = 0 . \tag{1}$$

This is the reduced Bernoulli equation. Here, the dot above the symbol denotes time differentiation, and $\sigma$ is the so-called deactivation factor of the earthquake source. The natural generalization of Eq. (1) is the Verhulst equation

$$\dot{n} = n(\gamma - \sigma n) . \tag{2}$$

Here $\gamma$ is the second phenomenological parameter of our theory.

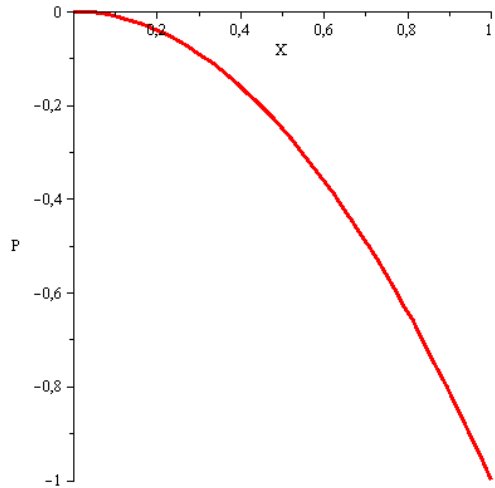

**Fig. 1.** Phase portrait of equation (1). The representative point moves from bottom to top along the phase trajectory with deceleration. The point (0, 0) corresponds to the equilibrium state.

Faraoni [7] proposed to introduce the phase plane $(n, \dot{n})$ of the dynamic system (1) simulating the activity of aftershocks. The corresponding phase portrait is shown in Figure 1. Instead of variables $n$, $\dot{n}$, we used dimensionless $X$, $P$ variables:

$$X = n / n_{\max} , \quad P = \left( n_{\infty} / \gamma n_{\max} \right) \dot{X} . \qquad (3)$$

Here $n_{\infty} = \gamma / \sigma$. The portrait consists of one phase trajectory that starts at point $(1, -1)$ and ends at point $(0, 0)$.

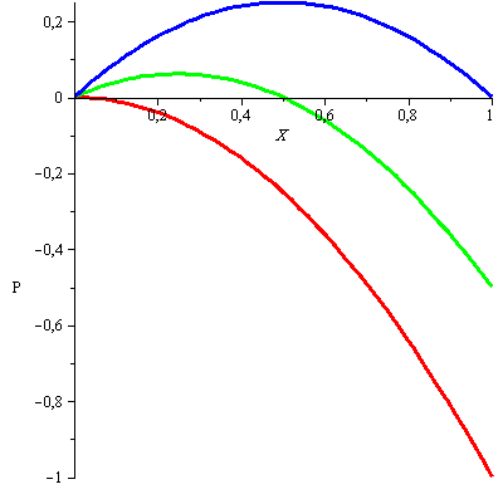

**Fig. 2.** Three phase trajectories of equation (2). The paths red, green and blue are plotted at $X_{\infty} = 0$, $0.5$ and $1$, respectively (see text).

Let's draw a phase portrait for equation (2). Figure 2 shows a family of phase trajectories constructed for different values of the parameter $X_{\infty} = n_{\infty} / n_{\max}$. The red, green and blue trajectories are plotted with $X_{\infty}$ values of $0$, $0.5$ and $1$, respectively. Equilibrium point $(0, 0)$ is stable at $X_{\infty} = 0$ and unstable at $X_{\infty} > 0$. Equilibrium point $(X_{\infty}, 0)$ is stable at any values of parameter $X_{\infty} > 0$. It can be shown that the velocity of motion of the imaging point along the phase trajectory asymptotically tends to zero with approaching $(X_{\infty}, 0)$. The segment of the trajectory located above the horizontal axis corresponds to the Verhulst logistic curve, widely known in biology, chemistry, sociology and other sciences (Figure 3). The segment located below the horizontal axis corresponds to the evolution of aftershocks (Figure 4).

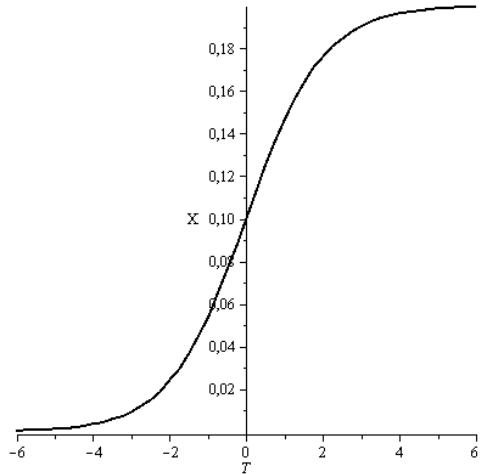

**Fig. 3.** Logistic curve at $X_\infty = 0.2$. Dimensionless time $T = \gamma t$ is plotted along the horizontal axis.

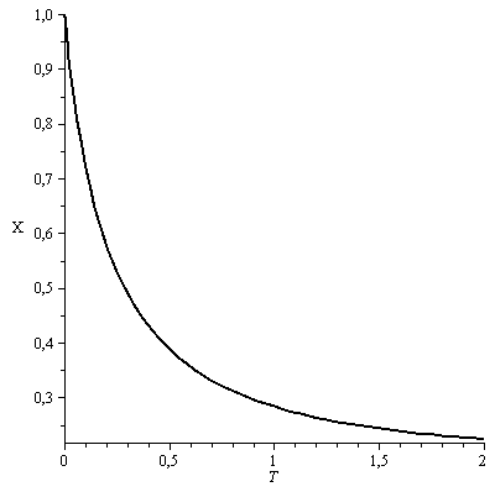

**Fig. 4.** Aftershocks curve at $X_\infty = 0.2$. Dimensionless time $T = \gamma t$ is plotted along the horizontal axis.

### 3. Omori epoch

The phase portrait in Figure 2 allows us to understand why the Omori hyperbolic law is most clearly manifested after strong earthquakes. Let us assume that there was a main shock with magnitude $M$. It is known that the activity of aftershocks, generally speaking, is the higher, the higher $M$. Accordingly, for a sufficiently large value of $M$, it is more likely that the initial condition $n_0 = n(0)$ in the Cauchy problem for the evolution equation (2) will be much more than the background value $n_\infty$. For $n_0 \gg n_\infty$, the term $\gamma n$ in Eq. (2) can be neglected at the initial stage of the evolution of aftershocks. In other words, when $n_0 \gg n_\infty$, the initial stage of evolution is approximately described by the truncated equation (1), the solution of which is completely equivalent to the Omori hyperbolic law (up to notation). Let's call the initial stage of the evolution of aftershocks the *Omori epoch*. The theory predicts that, other things being equal, the duration of the Omori epoch is the longer, the higher the value of $n_0$ and the lower the value of $n_\infty$.

The question arises whether the Omori epoch really exists, and how it ends. From a mathematical point of view, the Omori epoch smoothly passes into the final stage of evolution, which obviously ends with an asymptotic approximation of $n(t)$ from above to the equilibrium value of $n_\infty$ (see Figures 2 and 4). However, it turns out that in reality, evolution usually proceeds differently. Let's dwell on this issue in more detail.

Omori epoch is characterized by the fact that $\sigma$ is independent of time. The value of $\sigma$ can be determined experimentally as follows [2, 8]. Let's introduce an auxiliary variable

$$g = (n_0 - n)/nn_0,$$ (4)

and perform smoothing: $g \to \langle g \rangle$. In the Omori epoch, $\langle g \rangle$ is a linearly increasing function of time, and

$$\sigma = \frac{d}{dt}\langle g \rangle.$$ ( 5)

Note that the equation $d\langle g \rangle / dt = \sigma$, obtained by inversion of $n \to g = 1/n$, represents Omori's law in its simplest form.

Experience has shown that the linear growth of $\langle g \rangle$ is actually observed at the first stage of evolution [8]. In other words, the Omori epoch ($\sigma = \text{const}$) really exists. However, the Omori epoch usually ends with a far from smooth monotonic transition to the background, as the theory prescribes. Long before the inequality $n \gg n_\infty$ is violated, the linear growth of $\langle g \rangle$ is modified. The $\sigma(t)$ function calculated by formula (5) begins to experience a complex variation, and sometimes even becomes negative [8, 9].

We associate this phenomenon with the nonstationarity of the geological environment. Nonstationarity accompanies the relaxation process of earthquake source after a strong earthquake. Within the framework of our theory, the nonstationarity of the source deactivation can be taken into account by writing the general solution of Eq. (1) in the form

$$n(t) = \frac{n_0}{1 + n_0 \tau(t)}.$$ (6)

Here

$$\tau = \int\limits_{t_1}^{t} \sigma\left(t'\right) dt' . \qquad (7)$$

Formula (6) generalizes the classical Omori formula [4] and goes over into it for $\sigma = \text{const}$. Thus, the main motivation for writing the Omori law in differential form (1) is that it allows one to take into account the possible dependence of $\sigma$ on time.

## 4. Discussion and conclusion

Within the framework of our proposed phenomenological approach to the study of the evolution of aftershocks, we consider the shortened Bernoulli equation (1) as an axiological reduction of the Omori law. The binom $\dot{n} + \sigma n^2$ represents the semantic core of Omori law. Due to its distinct clarity, it enters both the Verhulst equation (2), which expands the possibilities of describing evolution in time [10], and the Kolmogorov-Petrovsky-Piskunov equation, which simulates the space-time evolution of aftershocks [11]. The $\sigma n^2$ term takes into account in its simplest form the effects of self-action, leading to a change in the initial properties of the geological environment under the influence of dynamic processes of high intensity.

Important in phenomenology is Brentano's principle of intentionality [12]. In our specific case, this means that we want to perceive equations (1), (2) not only as a convenient tool for analyzing aftershocks, but also as equations that have a certain meaning. Until now, attempts to give a physical meaning to the deactivation coefficient have remained unsuccessful, but they should not be stopped. An attempt to interpret $\sigma$ by analogy with the charge recombination coefficient in a plasma [6] was apparently unsuccessful. Whether the occurrence of aftershocks resembles a peculiar manifestation of the fundamental principle of Le Chatelier-Brown, similar to the electromagnetic Lenz effect, the thermoelectric Peltier phenomenon, and so on? It is unclear whether thinking along these lines will help to interpret the deactivation factor. For now, we must consider $\sigma$ simply as a phenomenological parameter of elementary theory (1) describing the quadratic self-action of aftershocks.

In the previous section, we noted that the coefficient $\sigma$, estimated experimentally using formula (5), undergoes complex variations after the end of the Omori epoch. Moreover, sometimes the parameter $\sigma$ becomes negative for some time. Let's discuss this anomaly in more detail.

Let $t_1$ and $t_2$ denote the beginning and end of the time interval at which $\sigma$ is less than zero. We redefine the time $\tau$ as follows

$$\tau\left(t\right) = \int\limits_{t_1}^{t} \left|\sigma\left(t'\right)\right| dt' , \qquad (8)$$

with $t_1 \leq t \leq t_2$. Then instead of (4) we will have

$$n(t) = \frac{1}{\tau_* - \tau(t)}, \qquad (9)$$

where $\tau_* = 1/n(t_1)$. Time $\tau$ is a monotonically increasing function of time $t$. If $\tau(t_2) < \tau_*$, then the function $n(t)$ remains regular, but on the interval $[t_1, t_2]$ there will be some increase in the frequency of tremors, reminiscent of the so-called swarm of earthquakes.

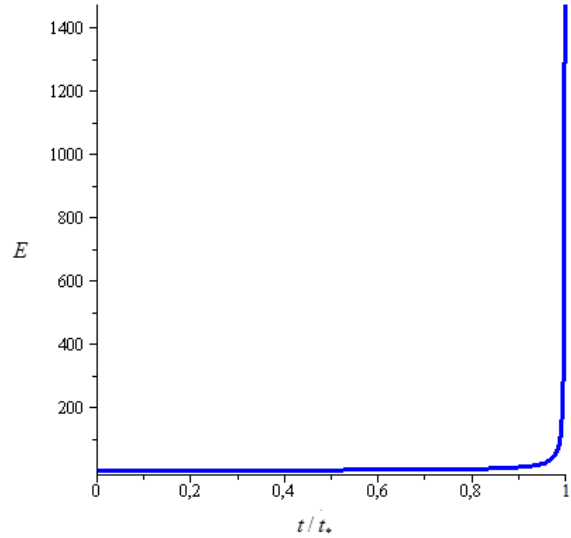

**Fig. 5.** Tectonic explosion when deactivation factor is negative. The vertical axis is energy in arbitrary units.

The case of the opposite inequality ($\tau(t_2) > \tau_*$) is more interesting. It follows from formula (9) that in this case on the interval $[t_1, t_2]$ a singularity of the function $n(t)$ arises at some time instant $t_*$ [10]. This behavior of a dynamical system is usually called explosive instability. In this regard, in [3] it was suggested that if, as an unknown function in the evolution equation, we choose not the frequency of tremors, but their energy $E$, then the explosive instability can be hypothetically presented as an image of the main shock of earthquake. Accordingly, in Figure 5 the vertical axis is energy $E$ (in arbitrary units). Thus, the theory points to the hypothetical possibility of a geotectonic explosion in the form of the main shock of an earthquake if the sign of the deactivation coefficient becomes negative. However, the reason for the possible change in sign of $\sigma$ is still completely unclear to us.

One circumstance gives rise to some doubt about the adequacy of the proposed scenario for the occurrence of the main shock in the form of a geotectonic explosion. On experience the main shock quite often occurs abruptly, without a noticeable increase in foreshock activity. Meanwhile, such an enhancement is definitely predicted by solution (9). Apparently, our theory is applicable to describe aftershocks and, possibly, earthquake swarms. To simulate the dynamic processes leading to the formation of the main shock, a radical modification of the theory is required. One of the search directions consists in the interpretation of the quantity $\sigma$ as a dynamic variable and the selection of an appropriate phenomenological equation that admits a rigid regime of self-excitation of the geological environment. In this case, the phase portrait of aftershocks on the $(n,\sigma)$ plane may turn out to be more interesting than the one that we presented on the $(n,\dot{n})$ plane. Moreover, through $\sigma$ we may be able to more naturally than it was done earlier [1, 2] to express the effect on the source of endogenous and exogenous triggers and to penetrate deeper into the question of self-oscillations of the Earth [14].

The incompleteness of our phenomenological theory also manifests itself in assessing the energy balance of processes in the earthquake source after the formation of the main rupture. Our theory belongs to the class of so-called "toy models". This will allow us to visualize the essence of the question.

Before the main shock, a large but finite elastic energy $E_0$ was stored in the volume of the source. The excitation of aftershocks after the main shock is a manifestation of the activity of one of the most important channels for the accumulated energy sink. It is quite understandable that the total energy of the aftershocks

$$E = \lim \int_0^t W(t)\ dt, \quad t \to \infty, \tag{10}$$

should be finite. Here $W$ is the power of the aftershock flow. Meanwhile, it follows from Eq. (1) that with a reasonable lower bound on the magnitude of aftershocks, integral (10) diverges logarithmically at the upper limit. This means that equation (1), which at $\sigma = \text{const}$ is completely equivalent to the classical Omori law, from the formal point of view, contradicts the energy conservation law. If $\sigma \neq \text{const}$, as is assumed in our theory, then the contradiction can be removed. It is sufficient to assume that at $t \to \infty$, the focal deactivation coefficient increases with time no slower than $\ln t$.

The situation is not the best with equation (2) at $\sigma = \text{const}$. At the same time, for $\sigma = \sigma(t)$ the divergence of integral (10) is easily eliminated by the standard cutoff procedure.

Thus, it is undesirable to use the Omori formula [4], which at $\sigma = \text{const}$ in the asymptotics gives the $n \sim (\sigma t)^{-1}$, since this, apparently, leads to the divergence of the integral (10). The well-known Utsu formula [15] is also unacceptable in this respect, since it leads to infinite energy at $p < 1$. (Here $p$ is the Hirano-Utsu parameter, see, for example, [1].) In contrast, the formula (6) does not contain a contradiction. It allows the regularization when calculating the energy integral (10). It would seem that formula (6) solves the problem, and it really is, but only in practical terms. Formula (6) is convenient to use when processing observation data, but this is not enough from a theoretical point of view. It is necessary to continue the search for a differential model describing the evolution of $\sigma(t)$.

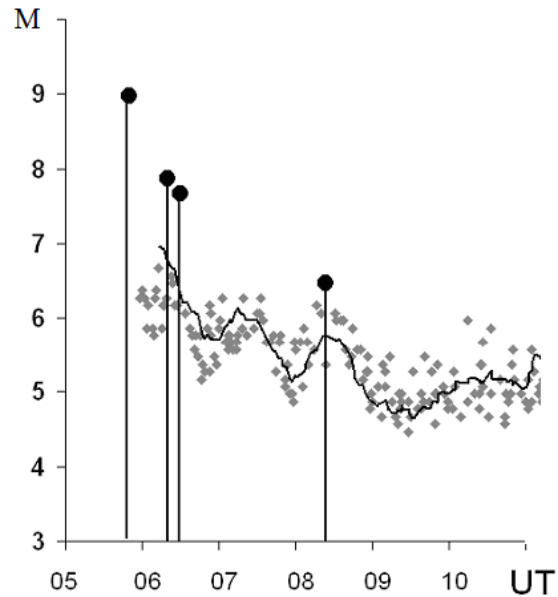

**Fig. 6.** The time distribution of magnitudes of the main shock ( $M = 9$ ) and aftershocks of the Tohoku earthquake, March 11, 2011.   Three strongest aftershocks are marked with black circles. The thin line approximates a cloud of weaker aftershocks [13].

In conclusion, let us briefly discuss the law of Bath, which was mentioned in the Introduction. Unlike Omori law, the Bath law is poor in mathematical content. Therefore, it is difficult to analyze it within the framework of phenomenological theory. Nevertheless, let us consider Figure 6 [13]. It represents the Tohoku earthquake. The main shock with a magnitude of M = 9.0 occurred on March 11, 2011 off the eastern coast of Honshu Island at 05:46 GMT. The strongest aftershock with a magnitude of M = 7.9 occurred at 06:16. The difference between the magnitudes is $\triangle M = 1.1$. We can say that formally the Bath law is fulfilled in this case. However, the substantial sense of this conclusion is not clear. Surprising is the too short time interval between the main shock and the strongest aftershock (30 min). It is not clear why three strong aftershocks (they are marked with black circles) do not fit into the general picture of the dynamics of aftershocks shown in the figure. It seems to us that the question of the strongest aftershocks in general, and of Bath law in particular, requires further careful study in experiment.

*Acknowledgments*. I express my sincere gratitude to B.I. Klain for a discussion of theoretical issues. Thanks to A.D. Zavyalov and O.D. Zotov for discussion of experimental aspects of the problem. I am grateful to F.Z. Feygin and A.S. Potapov for their interest in the work. And I am grateful to V.

Faraoni for his paper [7], which stimulated the writing of the submitted paper. The work was carried out according to the plan of state assignments of IPhE RAS.